\documentclass[11pt,a4paper]{article}
\usepackage{moriond,epsfig}
\usepackage{amsmath,amssymb}

\def\Journal#1#2#3#4{{#1} {\bf #2}, #3 (#4)}

\def\NIM{\em Nucl. Instrum. Methods}

\def\be{\begin{equation}}
\def\ee{\end{equation}}
\def\bea{\begin{eqnarray}}
\def\eea{\end{eqnarray}}

\newcommand{\jpsi}{\ensuremath{{J/\psi}}}
\newcommand{\D}{\ensuremath{{\mathrm d}}}

\begin{document}
\pagestyle{plain}
\vspace*{4cm}
\title{MEASUREMENT OF THE J/$\psi$ INCLUSIVE PRODUCTION CROSS-SECTION IN PP COLLISIONS AT $\sqrt{s}$=7\,TeV WITH ALICE AT THE LHC}

\author{ J. WIECHULA for the ALICE COLLABORATION }

\address{Physikalisches Institut der Universit\"at T\"ubingen, Auf der Morgenstelle 14,\\
72076 T\"ubingen, Germany}

\maketitle
\abstracts{
ALICE measures the \jpsi\ production at mid-rapidity ($|y|<0.9$) in the
di-electron decay channel as well as at forward rapidity ($2.5<y<4.0$) in the
di-muon decay channel. In both channels the acceptance goes down to zero transverse
momentum.
We present the rapidity dependence of the inclusive $J/\psi$
production cross-section and transverse momentum spectra.
}

\section{Introduction}
The measurement of the \jpsi\ production cross-section in pp collisions is
crucial for testing QCD models of quarkonium  production in the new
energy regime provided by the LHC.
Several theoretical models~\cite{th1,th2} have been proposed to describe the \jpsi\ production. 
However none of them consistently describes the production cross-section, the transverse momentum and rapidity dependences and the polarisation. 
Measurements of these variables are therefore essential to help understanding the underlying production mechanisms.

The main focus of ALICE~\cite{alice} is to study the deconfined hot and dense QCD phase which is expected to be formed in heavy-ion collisions. The \jpsi\ is an essential observable for this state of matter and the cross-section measurement in pp collisions is important as a reference.

ALICE measures the \jpsi\ production in two rapidity windows: 
at mid-rapidity ($|y|<0.9$) as well as at forward rapidity ($ 2.5 < y < 4 $).
We present first results of the \jpsi\ production cross-section as well as its $p_T$ and rapidity dependence in pp collisions at $\sqrt{s} = 7$\,TeV~\cite{jpsi}. The data were not corrected for feed-down contributions from other charmonium states ($\chi_c$, $\psi'$) or B-hadron decays (inclusive production).

\section{$J/\psi$ measurement with ALICE}
ALICE is a general purpose heavy-ion experiment. 
It consists of a central part with a pseudo rapidity coverage of $|\eta|<0.9$ and full acceptance in $\phi$ and a muon spectrometer placed at forward rapidity ($ -4 < \eta < -2.5 $). \jpsi\ production is measured in both rapidity regions, at mid-rapidity in the di-electron and at forward rapidity in the di-muon decay channel

The main detectors used for the analysis at mid-rapidity are the Inner Tracking System~\cite{alice,its} (ITS) and the Time Projection Chamber~\cite{tpc} (TPC). The ITS, placed at radii between 3.9\,cm and 43\,cm, consists of six cylindrical layers of silicon detectors, equipped with silicon pixel, silicon drift and silicon strip technology, two layers each. The main purpose of the ITS is to provide the reconstruction of the primary collision vertex as well as secondary vertices from decays. In addition it improves the momentum resolution of the TPC.
The main tracking device in ALICE is a cylindrical TPC. It has a length of 5 m and reaches from 85 cm to 247 cm in radial direction.
The TPC provides particle identification (PID) via the measurement of the specific energy loss ($\D E/\D x$) of particles in the detector gas. Due to the excellent $\D E/\D x$ resolution of $\sim 5.5\,\%$ PID could be performed for electrons with $1 \lessapprox p \lessapprox 6$\,GeV/$c$ using TPC information only.

The muon spectrometer consists of a 3\,T$\cdot$m dipole magnet, 5 tracking stations made of two planes of Cathode Pad Chambers each, a set of muon filters and a trigger system. Particles emerging from the collision in the forward direction have to traverse a front absorber with $10\,\lambda_I$, removing most hadrons and electrons. The remaining particles are reconstructed in the tracking system with an intrinsic position resolution of about 70\,$\mu$m in the bending direction.
An iron wall (7.2\,$\lambda_I$) is placed between the last tracking station and the muon trigger system.
The front absorber combined with the muon filter stop muons with momenta less than 4 GeV/$c$.
The trigger consists of 2 stations with 2 planes of Resistive Plate Chambers achieving a time resolution of $\sim 2$\,ns.
An additional absorber, placed around the beam pipe over the full length of the muon arm, protects the detector from secondaries produced in the beam pipe.

Data were collected using a minimum bias (MB) trigger condition, defined as the logical OR of at least one fired chip in the pixel detector and a signal in one out of two scintillator arrays which are placed in forward direction on either side of the interaction region. 
A muon trigger is required additionally in case of the di-muon analysis ($\mu$-MB condition).

For the analysis in the di-electron channel a total of $2.4\cdot 10^8$ MB events were analysed, corresponding to an integrated luminosity of $L_{\mathrm int} = 3.9$\,nb$^{-1}$.
Events are required to have a reconstructed vertex with a $z$ position within $|z_{vtx}|<10$\,cm from the nominal interaction point. Several cuts are applied on the level of the single track. Tracks are required to be in the detector acceptance ($|\eta|<0.9$) and have a transverse momentum larger than 1\,GeV/$c$. In addition the tracks have to fulfil certain reconstruction quality criteria. 
They need to be well defined in the ITS and TPC, the $\chi^2$ per cluster needs to be less than 4, at least 70 out of 159 clusters need to be attached in the TPC and a hit in at least one of the first two layers of the ITS is required. 
The latter helps to reject electrons from photon conversions. 
Finally electrons are identified by cutting on the $\D E/\D x$ signal of the TPC in terms of number of sigma, where $\sigma$ is the width of the gaussian response of the TPC ($\D E/\D x$-resolution).
The requirements are to be within $\pm 3 \sigma$ from the electron expectation and more than $3\sigma$ ($3.5\sigma$) away from the proton (pion) expectation.

\begin{figure}
\psfig{figure=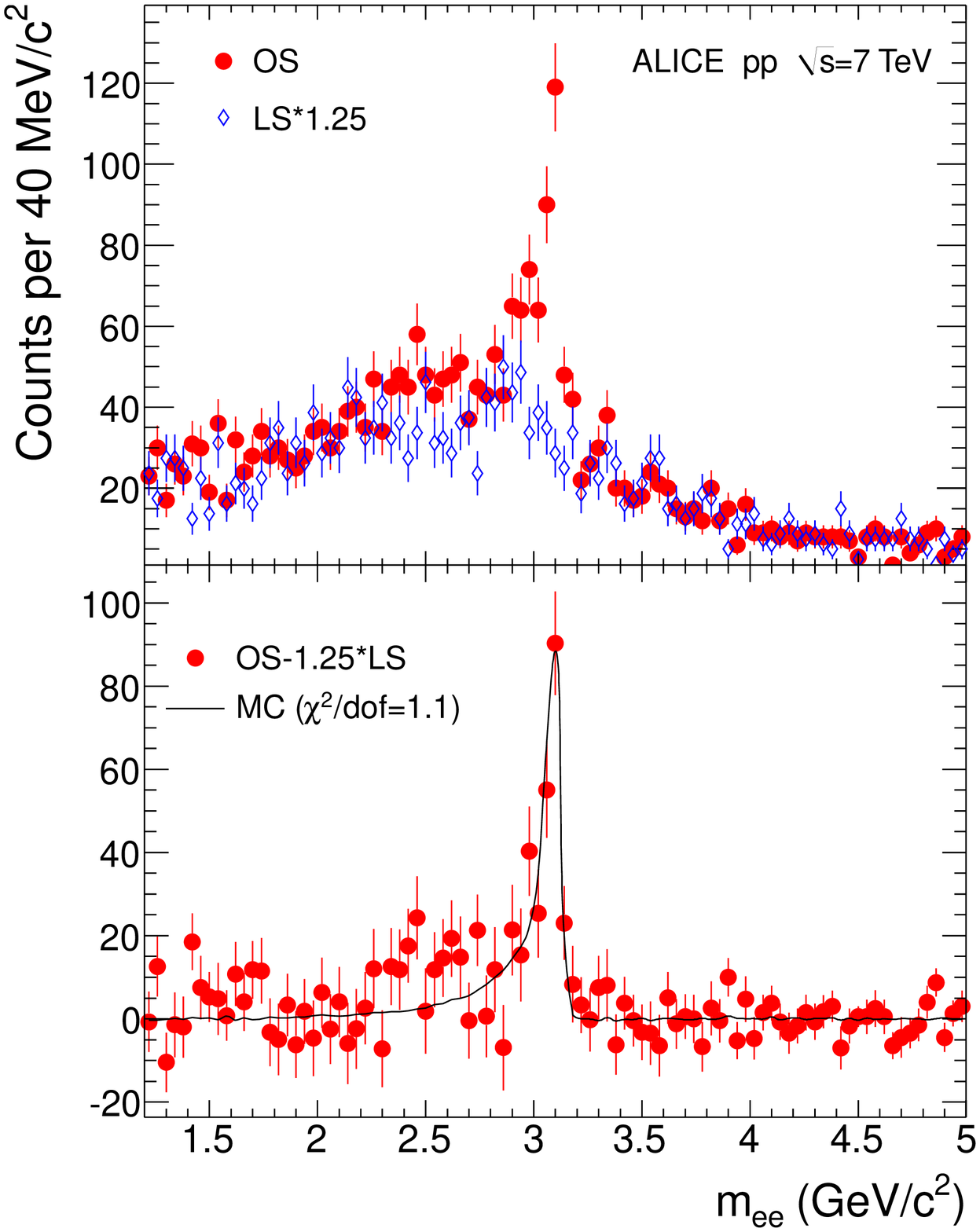,width=5.8cm}\hspace{1cm}\psfig{figure=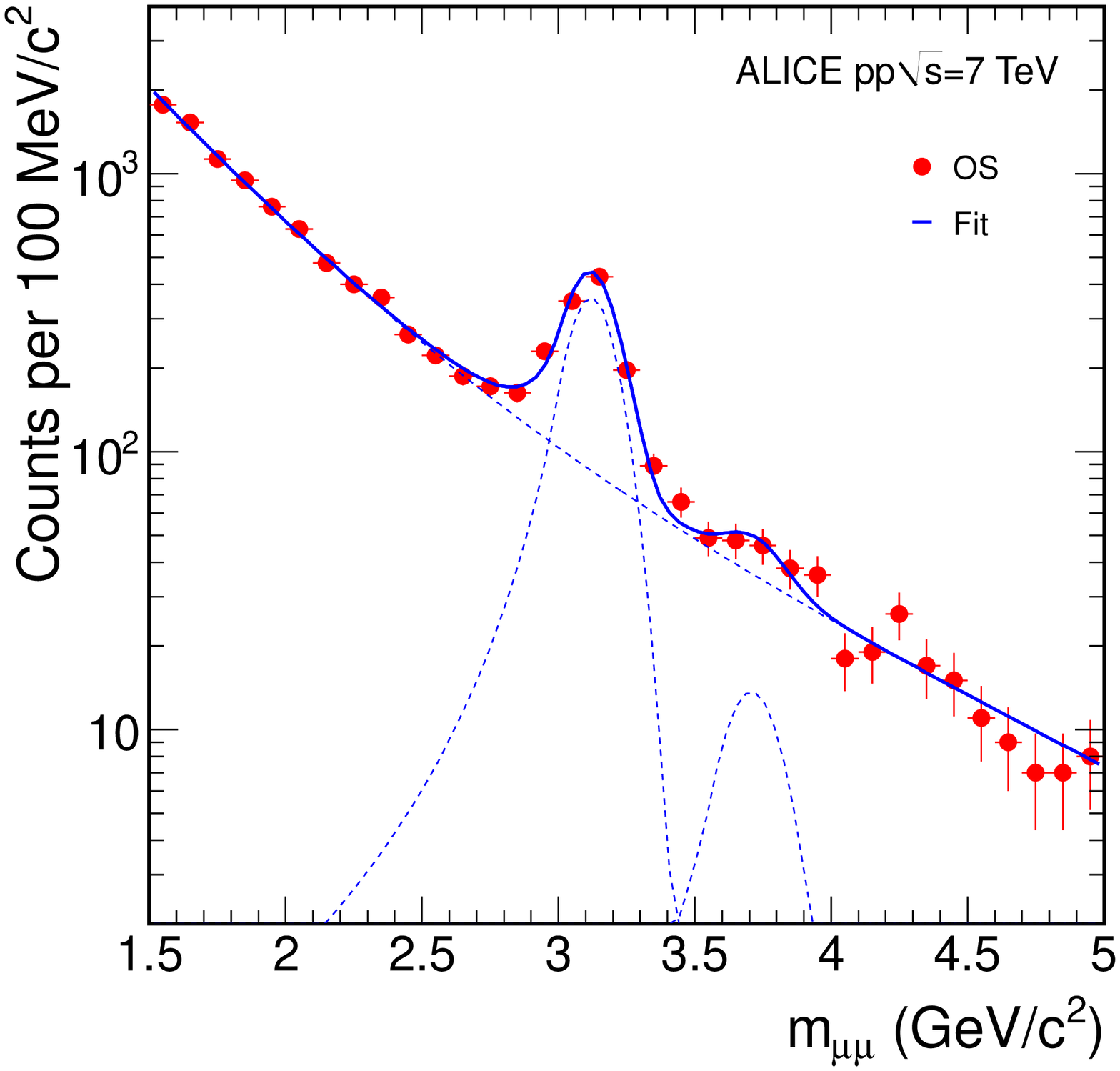,width=5.8cm}
\caption{Invariant mass spectra for the di-electron (left) and di-muon (right) decay channel~\protect\cite{jpsi}.}
\label{fig:minv}
\end{figure}

The obtained invariant mass spectrum of the opposite-sign (OS) di-electron pairs is shown in Fig.~\ref{fig:minv} (left, top panel, red points). 
The remaining background is described by combining like-sign (LS) di-electron pairs as $N^{++}+N^{--}$ and scaling the LS spectrum to match the integral of the OS spectrum in the invariant mass range of $3.2 - 5$\,GeV/$c^2$ (Fig.~\ref{fig:minv}, left, top panel, blue points). The arithmetic mean was preferred over the geometric mean in order to remove the bias that would occur in the scaling due to bins with zero entries in the LS spectrum.
After background subtraction the number of \jpsi\ candidates is extracted by bin counting in the invariant mass range $2.92-3.16$\,GeV/$c^2$. 
This yields N$_\jpsi$ = 249 $\pm$ 27 (stat.) $\pm$ 20 (syst.). The systematic uncertainties are described below.

The analysis in the di-muon channel was performed on a data sample of $1.9 \cdot 10^8$ $\mu$-MB events ($L_{\mathrm{int}}=15.6$\,nb$^{-1}$). For further analysis only events were selected for which at least one muon candidate has a match in the muon trigger chambers. This significantly reduces the background of hadrons which are produced in the absorber. Further selection criteria are a reconstructed vertex in the ITS, the rejection of muons emitted under very small angles, which cross a significant fraction of the beam pipe and requiring the di-muon pair to be in the detector acceptance ($2.5 < y < 4$) in order to minimise edge effects.

Applying all selection criteria $1.75 \cdot 10^5$ OS muon pairs are found. Figure~\ref{fig:minv} (right) shows the invariant mass spectrum of OS di-muon pairs of a subset of data. The signal is extracted by simultaneously fitting a Crystal Ball function to describe the signal shape and two exponentials for the background. By integrating the Crystal Ball function over the full mass range, the extracted \jpsi\ yield of the complete analysed statistics is N$_\jpsi$ = 1924 $\pm$ 77(stat.) $\pm$ 144(syst.).

The systematic uncertainties were obtained considering various sources: signal extraction, acceptance effects due to the $p_T$ and rapidity distributions used as input for the MC studies, muon trigger efficiency, reconstruction efficiency, error on the luminosity measurement and the uncertainty of the branching ratio. 
The total uncertainty is 12.6\,\% in the di-muon and 16.1\,\% in the di-electron channel. However, the largest uncertainty results from the unknown polarisation of the \jpsi\ due to its influence on the acceptance corrections. To quantify the effect, the impact of fully transverse ($\lambda=1$) and fully longitudinal ($\lambda = -1$) polarisation for the case of the helicity (HE) as well as the Collins-Soper (CS) reference frame was investigated. Maximum variations between -15\,\% and +32\,\% were estimated and will be quoted separately. For details see~\cite{jpsi}.

\section{Results}
The production cross-section is determined by normalising the efficiency and acceptance corrected signal ($N_\jpsi^{corr} = N_\jpsi/(A \times \epsilon)$) to the integrated luminosity or the cross-section of a reference process. For this analysis the minimum-bias cross-section itself was chosen as a reference. Thus the \jpsi\ cross-section is given by\vspace{-.3ex}
\begin{equation}
\notag
\sigma_\jpsi = \frac{N_\jpsi^{corr}}{BR(\jpsi \rightarrow l^+l^-)} \times \frac{\sigma_{MB}}{N_{MB}},
\end{equation}
where $BR(\jpsi \rightarrow l^+l^-) = (5.94 \pm 0.06)\%$ is the branching ratio of \jpsi\ to di-leptons, $N_{MB}$ is the number of analysed minimum bias events and $\sigma_{MB} = 62.3 \pm 0.4$ (stat.) $\pm 4.3$ (syst.)\,mb is the minimum bias cross-section, which was derived from a Van-der-Meer scan~\cite{sigmaMB}. 
The acceptance times efficiency value ($A \times \epsilon$) is 10.0\,\% for the di-electron, 32.9\,\% for the di-muon decay channel.

The measured integrated cross-sections for the two rapidity ranges are:\\
$\sigma_\jpsi (|y| < 0.9) = 10.7 \pm 1.2$ (stat.) $\pm 1.7$ (syst.) + 1.6 ($\lambda_{HE}$ = 1) - 2.3 ($\lambda_{HE}$ = -1) $\mathrm \mu$b and \\
$\sigma_\jpsi (2.5 < y < 4) = 6.31 \pm 0.25$ (stat.) $\pm 0.80$ (syst.) $+ 0.95 (\lambda_{CS} = 1) - 1.96 (\lambda_{CS} = -1)$ $\mu$b.

\begin{figure}
 \centering
\psfig{figure=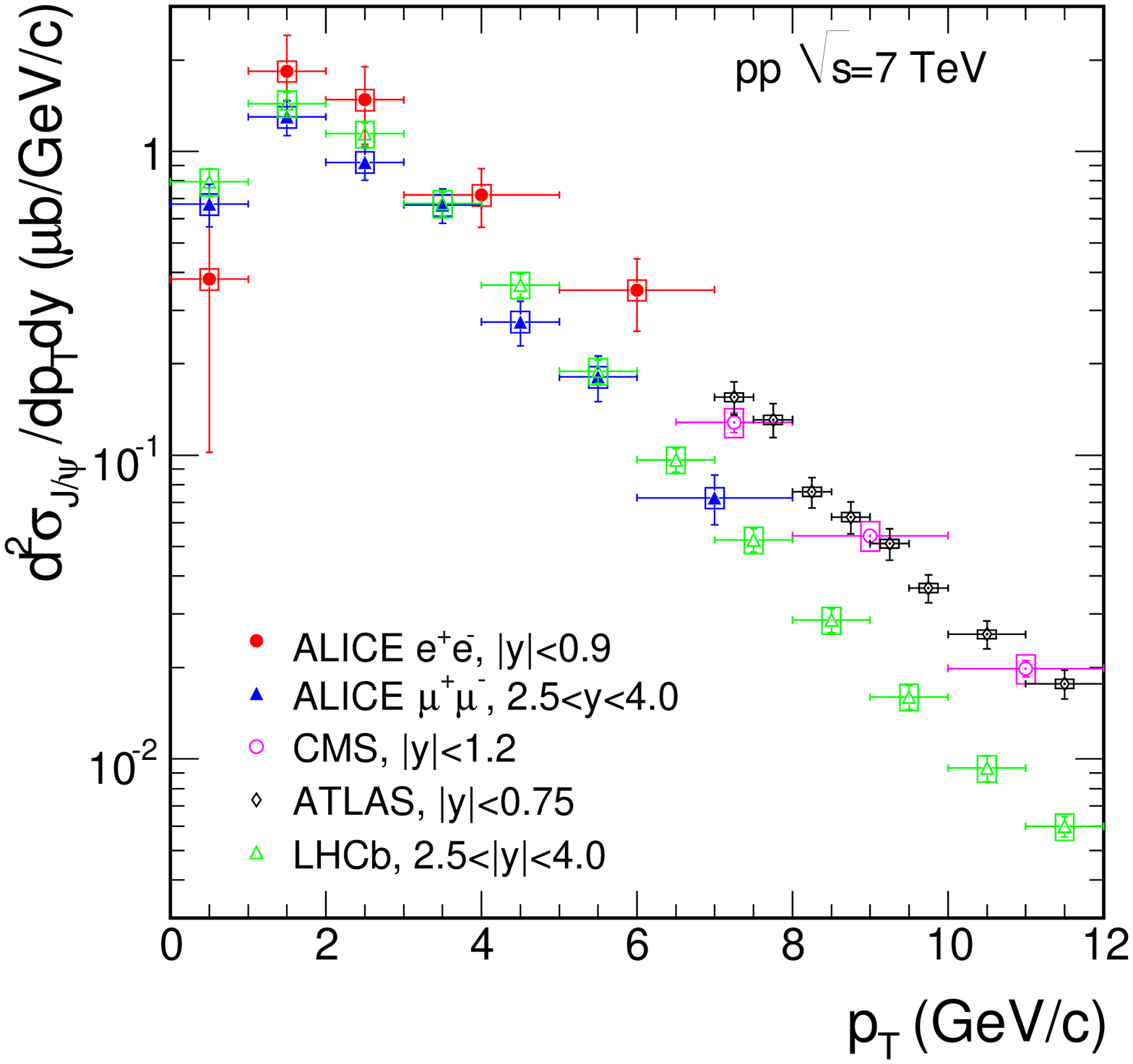,width=6cm}\hspace{1cm}\psfig{figure=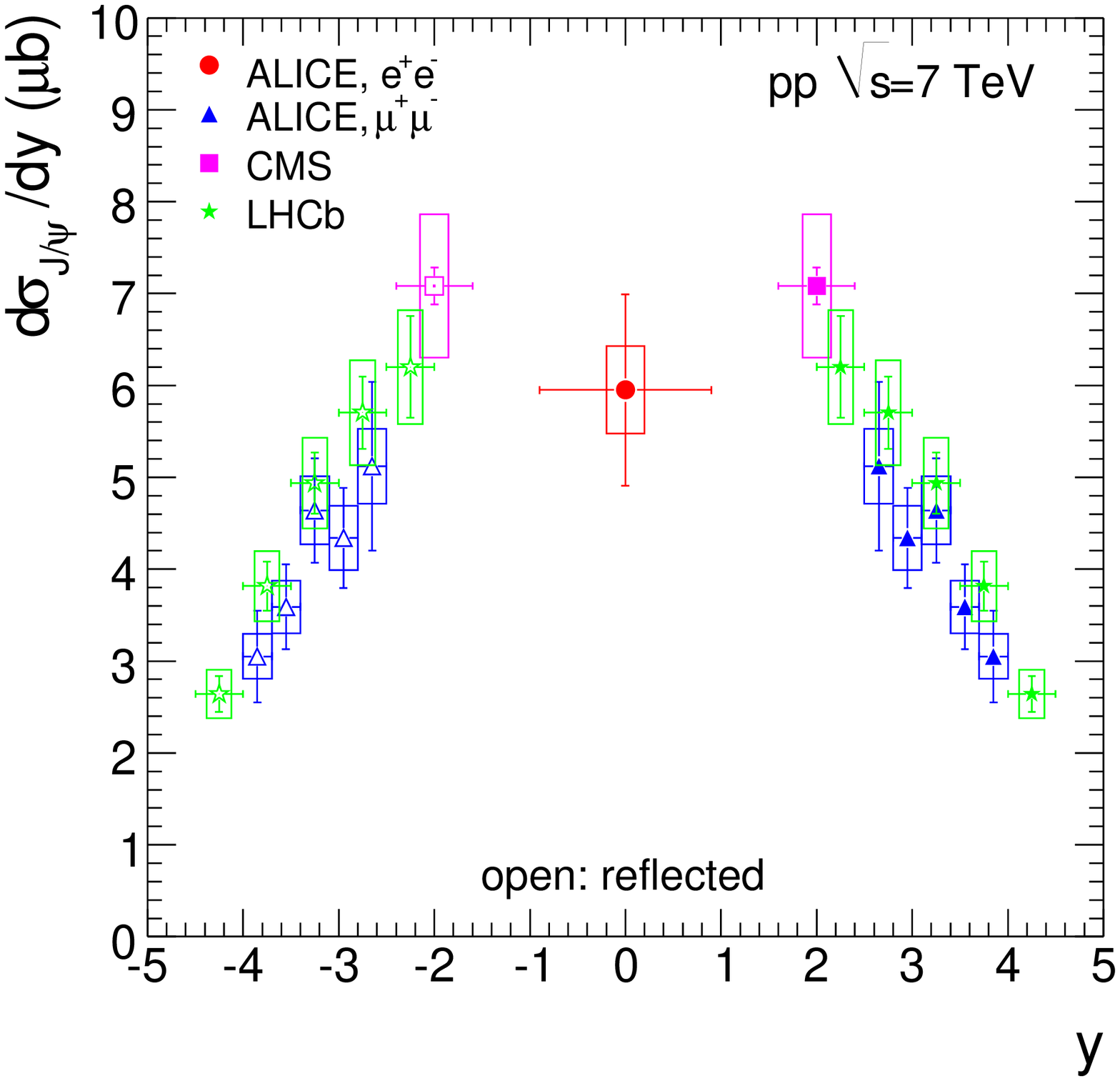,width=6cm}
\caption{Differential \jpsi\ production cross-sections as a function of $p_T$ (left) and rapidity (right)~\protect\cite{jpsi}.}
\label{fig:radish}
\end{figure}

In addition for both rapidity ranges the differential cross-sections $\D^2\sigma_{\jpsi}/\D p_T\D y$ and $\D\sigma_{\jpsi}/\D y$ ($p_T > 0$) were determined. The spectra are presented in Fig.~\ref{fig:radish} and compared with results from ATLAS~\cite{ATLAS} ($p_T>7$\,GeV/$c$, $|y|<0.75$), CMS~\cite{CMS} ($p_T>6.5$\,GeV/$c$, $|y|<1.2$) and LHCb~\cite{LHCb} ($p_T>0$, $2.5<y<4$). All results are compatible. The ALICE measurement at mid-rapidity extends down to $p_T=0$ an thus is complementary to those of ATLAS and CMS.

\section{Summary and Outlook}\label{subsec:fig}
We have presented first results on the \jpsi\ inclusive production cross-section measured with the ALICE detector system. In addition the $p_T$-differential cross-section and the rapidity dependence were shown and compared to the other LHC experiments. ALICE is the only experiment at LHC measuring \jpsi\ at mid-rapidity down to zero transverse momentum.

With increased statistics and a different PID strategy, the large statistical as well as systematic uncertainty of the measurement at mid-rapidity will be reduced significantly. To provide a better electron identification other detectors will be used in addition to the TPC. Under investigation are the Time Of Flight detector, to extend the electron identification towards lower momenta ($p \lessapprox 1$\,GeV/$c$) as well as the Transition Radiation Detector to allow a better electron to pion separation at higher momenta ($p \gtrapprox 2$\,GeV/$c$).

Ongoing analyses in both decay channels are the measurement of the feed-down from B-hadron decays, the analysis of pp data at $\sqrt{s}=2.76$\,TeV and the multiplicity dependence of the \jpsi\ production in pp collisions.
In addition the measurement of the polarisation will help to put limits on the largest contribution to the systematic uncertainty.

In parallel to the analysis of pp collision, the data from the Pb--Pb run at $\sqrt{s_{NN}} = 2.76$\,TeV of November/December 2010 are being analysed. First results were presented at the Quark Matter 2011 conference~\cite{QM}.

\end{document}